# 25

# An Online Environment for Democratic Deliberation: Motivations, Principles, and Design

TODD DAVIES, BRENDAN O'CONNOR , ALEX ANGIOLILLO COCHRAN, JONATHAN J. EFFRAT, ANDREW PARKER, BENJAMIN NEWMAN, AND AARON TAM

## 1  Introduction

We have created a platform for online deliberation called *Deme* (which rhymes with 'team'). Deme is designed to allow groups of people to engage in collaborative drafting, focused discussion, and decision making using the Internet.

The Deme project has evolved greatly from its beginning in 2003. This chapter outlines the thinking behind Deme's initial design: our motivations for creating it, the principles that guided its construction, and its most important design features. The version of Deme described here was written in PHP and was deployed in 2004 and used by several groups (including organizers of the 2005 Online Deliberation Conference). Other papers describe later developments in the Deme project (see Davies et al. 2005, 2008; Davies and Mintz 2009).

Demes were the divisions or townships of ancient Attica (from the Greek word *demos*—the populace). In ecology, a deme is a local population of closely related plants or animals, and in modern Greece a deme is a commune (OED, 1989). The name was chosen to reflect our focus on providing an online tool for small to medium-sized groups that (1) have a substantial face-to-face existence that predates or is independent of any interaction on the Internet, (2) are geographically limited so that all members can meet each other face to face, and (3) have difficulty meeting face to face as much as they need or would like to. Examples of such groups include neighborhood associations, places of worship, community interest groups, university groups (e.g., dormitories), and coalitions of activists.

Targeting this type of group suggested a distinct set of design criteria from those that govern groupware for 'virtual' (Internet-based) groups, businesses, or large organizations. The decline in participation, within the U.S., in small, community-based civil society groups such as the ones we are targeting has received considerable attention from political scientists and sociologists (see Putnam 2000; Skocpol 2003).

## 2   Background

In January of 2002, students and faculty affiliated with the Symbolic Systems Program at Stanford University began a consultative partnership with staff of the newly forming East Palo Alto Community Network. East Palo Alto is a vibrant, low-income, multi-ethnic, and multi-lingual community of 29,506 residents (U.S. Census 2000), located three miles from the Stanford campus. The East Palo Alto Community Network[1] comprises a community website or 'portal' (EPA.Net) and technology access points ('TAPs'—public computer clusters located throughout the city).

Over the first year (2002) of this partnership, which became the Partnership for Internet Equity and Community Engagement (PIECE), projects included studies of how the needs of the area's diverse groups related to the Internet, and of the realized and unrealized role of Internet tools in improving civic engagement.[2] In the second year (2003), we focused on (1) outreach to the community, (2) follow-up data collection to assess the impact of the community website one year after its launch, and (3) designing a tool for online deliberation, which is the topic of this chapter.

## 3   Motivations

In an earlier paper (Davies et al. 2002), members of our team argued that East Palo Alto residents and community organizations could gain a great deal through the use of the Internet. This was one of the motivating principles behind the Community Network and other recent technology initiatives in East Palo Alto. Our research looked especially at barriers that keep resi-dents from knowing about, participating in, and influencing decisions that affect them, and at how Internet tools could reduce or eliminate those barriers.

---

[1] The community network has been funded primarily by grants from Hewlett Packard and the National Telecommunications and Information Administration's Technology Opportunities Program (TOP), with software donations from Microsoft.

[2] These and other projects, including Deme, are discussed on the PIECE website (http://piece.stanford.edu, last accessed January 18, 2009).

Our early research drew two broad conclusions concerning the use of Internet tools for enhancing democratic decision making in East Palo Alto:

- The ability to use computers and the Internet was distributed very unevenly within the community, and was especially absent among Spanish-speaking residents who do not speak English very well (68% of the Latino population, which is 59% of the city; U.S. Census 2002; Sywulka 2003). We refer to eliminating 'digital divides' as the goal of ***Internet equity***.
- When the ability to use the Internet is commonplace among members of a group, Internet communication can address many of the difficulties associated with democratic participation in East Palo Alto's organizations and the City Government, for that group of Internet users. Using both the existing community website (EPA.Net) and developing new networking tools appear necessary to best achieve the goal we refer to as ***community engagement***.

Much of the Community Network's expenditure and effort, and some of PIECE's work, has been aimed at improving Internet equity (the focus of the first conclusion) through, for example, providing hub computer access and training open to all residents, making the content and functionality of the EPA.Net website motivating and accessible (e.g., through community news coverage and automatic translation), and reaching out to community network users and potential users.

The present chapter primarily concerns the *second* conclusion. The PIECE team explored community engagement through both research and tool development. We began by attending several types of meetings, including those of advisory boards for nonprofit organizations, informational and feedback meetings open to community or neighborhood members, and official functions of the City Government, and by subscribing to both organizational and community email lists in East Palo Alto.

Through participant-observation, reading, and interviews, we found that most group decisions made in East Palo Alto occurred in face-to-face meetings, often involving volunteers or people who received little compensation for participating. Residents had, in many cases, very little free time (e.g., they worked double shifts or had long commutes to their jobs). There was a widespread perception that decisions were made by a handful of people who served on multiple committees, were well connected, and sometimes had their own agendas, and that groups were not empowered in proportion to their population. Although our observations generally indicated a high level of interest, effort, and public spiritedness among the city's leaders, this substantive reality was sometimes undermined by perceptions of procedural injustice (see Tyler 1988).

This situation is mirrored in many communities. We found a number of recurring community engagement difficulties that Internet tools might address (see also Davies et al. 2002):

1. *Attendance and representation*. When attending a face-to-face meeting is the only way to have input into a decision, many people are disenfranchised because they cannot attend, because of work or family obligations or other engagements, and this is likely to make attendees collectively less representative of all stakeholders.

2. *Meeting duration and frequency*. When meetings are not held very frequently (frequent meetings being difficult for everyone to attend), or when the time available for meetings is scarce, groups are less able to act in ways that are timely, or with adequate discussion.

3. *Communication between meetings*. When groups lack efficient means for communicating between meetings (e.g., if they do not have an email list, or if not everyone is on the list), meeting quality suffers because attendees are likely to be under prepared, or worse, they may not know the time/location of the next meeting.

4. *Available information during meetings*. When decisions are made in a setting where some or all attendees are unable to access information that may be relevant to a decision (e.g., a room with no computer or Internet connection, or the relevant experts not present), meeting quality suffers because attendees must rely on memory, common knowledge, or the word of others who may persuade them, rather than basing opinions on the best information.

5. *Communication between groups*. When groups' decisions affect each other (e.g., subcommittees, groups in coalition, or multiple stakeholders), traditional means of communication between them are often inadequate, leading to conflicts, duplicated effort, and uninformed decisions.

6. *Group records*. Groups making decisions in face-to-face meetings often have inadequate records of their own past deliberations and decisions when they meet, which can lead to disputes, conflicting decisions that must be revisited, and duplication of previous effort.

7. *Decision procedures*. Face-to-face meetings often lead to streamlined, time-saving procedures for making a decision, which may not fit the complexity of what the group must decide, or which may unduly empower the chair or agenda committee (e.g., presentation-sensitive procedures, voting that does not take into account relative preferences among multiple options, etc.).

8. *Transparency*. Face-to-face meetings are difficult to record or to broadcast, so that those who cannot be present are often left unable to know exactly what has happened. This can lead to mistrust, side dealing, and general disenfranchisement.

The above findings point to a clear role for Internet communication as either a supplement to, or in some cases a replacement for, face-to-face decision making. Many of the above observations would apply to more affluent communities, and we have observed them in many settings outside of East Palo Alto. But the difficulties posed by an almost exclusive reliance on face-to-face meetings are amplified in East Palo Alto because, in comparison to the more affluent residents of neighboring communities: (1) East Palo Alto's residents are more dependent on community resources; (2) they have more experience with being disenfranchised or otherwise being victimized (and are therefore more likely to break off trust relationships); and (3) they have fewer means to participate outside of public forums, which they may be unable to attend. Prior to EPA.Net, East Palo Alto did not have its own media (newspapers or a broadcast station).

Some of the challenges we have identified for community engagement in East Palo Alto were addressed through already existing features of the Community Network: e.g., getting organization members access to the Internet and email accounts, setting up email lists, collecting relevant information about groups and the city on the community website, and publicizing important meetings. But to address the above challenges fully requires a kind of groupware that did not appear to exist before we began this project.

## 4 Principles

The challenges listed above (1 through 8) led easily to the idea that Internet tools could improve group decision making, if the group's members each had regular Internet access. Attendance and participation would be easier because members would not have to travel to participate, and if the tool allowed asynchronous communication, members could participate at their convenience instead of needing all to be present at the same time. Discussion comments could be composed at a more leisurely pace and with more care, and the group would not be constrained by its announced meeting times and durations. Even if face-to-face meetings were to continue to be the primary setting for decisions, Internet communication could occur between meetings, and relevant outside information as well as communication with other groups could be more easily incorporated into discussions through linking. An online archive of the group's activities would make it less likely that the group would get bogged down due to a lack of collective memory, and, since the Internet can be used as a form of broadcasting, all stakeholders could follow what was happening in the proceedings of a group.

The observation that Internet communication could address challenges 1-8 was, however, just a starting point. How could the Internet best be used to address these difficulties, serving the general goal of enhancing the ability of group members and/or stakeholders to participate in decisions that affect them? We concluded that the design of a platform or toolset for groups that have a substantial non-Internet existence, should ideally satisfy four top-level criteria. The criteria take the form of outcome goals that we intended to be evaluated with respect to a particular group or set of groups.

The first criterion required that online interaction enhance, or at least not diminish the group's overall effectiveness, on- and offline. We called this the criterion of supportiveness.

> ***Supportiveness***. The platform should support the group overall, so that there is either an improvement or no decline in the ability of the group to meet the needs of its members or stakeholders.

The second criterion (comprehensiveness) expressed a desire to liberate the group from a dependence on having face-to-face meetings. While groups might still choose to meet face to face, eliminating the need to *rely* on face-to-face meetings would mean that there would no longer be an excuse for inner-circle, closed-door decision making, because no task would require it.

> ***Comprehensiveness***. The platform should allow the group to accomplish, in an online environment, all of the usual deliberative tasks associated with face-to-face meetings.

The third criterion expressed a desire to make decision making more participative relative to what occurs in face-to-face meetings.

> ***Participation***. The platform should maximize the number of desired participants in the group's deliberations, and minimize barriers to their participation.

Finally, the fourth criterion (quality) was aimed at the group's satisfaction with the process and substance of its decisions.

> ***Quality***. The platform should facilitate a subjective quality of interaction and decision making that meets or exceeds what the group achieves in face-to-face meetings.

Combining these four criteria with general principles of design yielded a richer set of design principles. These derived principles were closer to the level of actual design, and provided an outline of the functionality for our platform. In the subsections below, we discuss the design principles and goals (highlighted in *italics*) that we derived from each of the four outcome criteria listed above.

**Supporting the Group**

The criterion of supportiveness is analogous to Hippocrates' famous dictum 'do no harm'[3]. We interpreted supportiveness, in part, to mean that groups should have *autonomy* over the toolset that they use for deliberation, so that group members can determine as much as possible for themselves when and how to use online tools, how and whether to modify them, and what resources should be devoted to their maintenance. Inasmuch as tools can be made available as *free and open source* software, supportiveness does not seem consistent with a model that draws resources away from groups (e.g., monetary payments by the group that exceed or are not tied to fair compensation for labor and other costs), or that limits access to online tools for commercial purposes or to benefit the provider at the expense of groups. Open access to the code seems especially important for software that is going to be used for decision making (e.g., elections), where group members may worry whether they can trust the results.

Supportiveness also implies that online deliberation should not lead to reduced satisfaction with the group on the part of its members or stakeholders. The online platform should therefore *build in feedback and assessment* from group members, shared both within the group and with tool providers, at different stages during and after tool adoption.

As a guiding principle, a supportive platform should not take away capabilities that the group possessed before its adoption, but should *integrate with existing practices* as much as possible. If group members are using email as a group communication tool, for example, and want to continue doing so, supportiveness implies that any new platform should incorporate email usage where it can be accommodated, without also diminishing the effectiveness of the earlier practice (e.g., by maintaining the option to communicate with the group by email and not creating a separate interaction space that is unnecessarily inaccessible through email).

**Comprehensive Deliberation**

The criterion of comprehensiveness implies that we can map the usual activities of face-to-face deliberation onto the design of an online toolset. Meetings in organizations feature discussion that is typically *focused on*

---

[3] This appears not in the Hippocratic Oath itself but rather in Hippocrates' *Epidemics,* Bk. I, Sect. XI.: 'As to diseases, make a habit of two things—to help, or at least to do no harm' (http://www.geocities.com/everwild7/noharm.html, last accessed December 20, 2008).

*particular agenda items*. These items give structure to the meeting, and are usually discussed in some order. One type of agenda item is simply a topic of discussion, such as a question about which members of the group brainstorm or express their opinions. Discussion items are often well suited to traditional online forums (e.g., Web message boards or even listservs) because the topic can generally be specified simply (e.g., by posting a question). But organizations often must go beyond exchanges of opinion to *numerical polling or formal decisions,* in which some agreed upon procedure is applied, such as voting or testing for consensus. Furthermore, each group has its own procedures for decision making, and if an online platform is to provide comprehensive support for the group's deliberations, it must give the group *options for decision making procedures* that are sufficiently close to its offline practice.

A general design principle of *flexibility and customizability* derives from the goal of comprehensive deliberation support. This can also be applied to another important type of agenda item: the drafting of a document. Documents such as bylaws, flyers, press releases, and budgets, should ideally be expressions of a group's will. *Collaborative drafting* is a cumbersome process that often gets delegated to one or a few people who can meet face to face. But the power that is delegated in such cases can be considerable. Even if the whole group must ultimately approve a document, those who participate in drafting it in its earlier stages are likely to have disproportionate influence on its content. At a minimum, an online platform should support the same level of document collaboration as can occur in face-to-face meetings. At best it offers the possibility of exceeding that standard.

Documents (including nontextual material such as images and video recordings) can be objects of discussion in meetings both as part of collaborative drafting and as the centerpieces of debate (e.g., as evidence that bears on a decision). An important feature of face to face meetings, in contrast to the lists of messages that usually comprise online discussion, is that a document can be placed in the common view of a meeting's participants, by distributing copies or projecting it onto a screen, and oral discussion can center on the document through synchronizing references (such as: 'Everyone look at the paragraph beginning with "Maria said...".'). The importance of common views or WYSIWIS ('What you see is what I see') has been stressed from the early days of research on computer-supported cooperative work (see Stefik et al. 1987). The ability of meeting participants to function simultaneously in two discourse spaces—the document and the discussion, generally by applying separate perceptual modalities (visual and auditory), is a formidable advantage of face-to-face meetings that must somehow be captured in a fully online platform if the criterion of comprehensiveness is to be met, to allow *document-centered discussion* (Davies et al. 2008).

The structure of both civil society and government groups typically resembles a network of clusters, exhibiting relatively high levels of connectivity within groups (clusters), and low (though important) connectivity between groups. This argues for *each group having its own online space*, with the ability to close access for nonmembers, but also to establish lines of communication with other groups. Groups usually include subgroups such as committees, or they may segment meetings into different topics. These observations imply that each group should be able to create *separate online spaces for different subgroups or meeting topics*. Often, groups of representatives from different groups form coalitions, which implies that *meeting areas should be able to be linked across groups* as well.

Collaborative drafting, document-centered discussion, rich support for decision procedures, and hierarchical and network structuring of group meeting spaces are all cumbersome in standard message-list online environments. We therefore emphasized these in our design principles/goals for an online deliberation environment. There are many other activities associated with face-to-face group meetings that were well supported in groupware prior to 2003, such as *announcements*, the keeping of a *common calendar*, the *sharing of personal information* by group members, and *the ability to share files and links*. Since we assumed that groups would desire minimal inconvenience in moving between these capabilities, we inferred that they should be integrated with a deliberation toolset so that groups could have an all-purpose online space to call their own.

**Maximizing Participation**

The participation criterion has a number of consequences for the design of a deliberation platform. Maximizing the number of people who can participate implies that communication should be *asynchronous* so that group members can participate at their own convenience. The software should be *compatible and interoperable* with the widest possible range of server and user environments, so that those who might participate are not prevented from doing so for technical reasons.

Participation is affected by other factors that determine how comfortable group members feel using the platform, e.g., *familiarity of features, design simplicity* and *intuitiveness, accessibility to those with special needs, execution speed* and *robustness,* trustworthy *privacy protection,* and *secure communication*.

For those who can use an online deliberation tool, overall participation may be enhanced merely by this fact. A number of authors have noted the tendency of computer-mediated communication to equalize participation (see Kiesler et al. 1984; Price 2009). Of course, accessibility is key in realizing this potential.

**High Quality Deliberation**

The criterion of quality could be assessed subjectively, through the kind of *built-in feedback* referred to above under 'Supporting the Group'. There are also numerous principles that have been proposed for creating sound deliberation, such as the conditions of the 'ideal speech situation' defined by Habermas (1990; Horster 2001; see also Kiesler et al. 1984), and other theorists of 'deliberative democracy' (see Gutmann 1997). In general, enhancing decision quality seems to call for greater *structure around which discussion can take place*. Farnham et al. (2000) demonstrated that more structured discussion in a chat room (i.e. preauthored scripts) improves the ability of a group to come to consensus.

A full treatment of the theory of deliberation is beyond the scope of this paper, but it seems possible for an online platform to support good discourse practices through, for example, *built-in tutorials* and *models of practice*, as well as *features that encourage directed discussion* (e.g., encouragement to quote comments being responded to, when possible, rather than to paraphrase them; clear options for one-on-one replies when a more visible discussion is not justified, etc.).[4]

## 5 Design

Applying the above principles within what is technically and otherwise feasible for us led to the creation of Deme: an online environment for group deliberation. In this section and the next, we describe the early design of Deme and attempt to relate its features to the design principles and goals derived in the previous section. The early design of Deme was refined through a series of meetings with prototypical target groups: the Community Network staff in East Palo Alto, prospective users at Stanford, and a grassroots group of labor activists organizing a labor media/technology conference. These groups provided valuable input to the design, and Deme's features reflected their comments.[5]

We organized Deme around *group spaces*: subsites that were each devoted to a particular group. A *group* was assumed to be either a well-defined set or a looser cluster of individuals who identified themselves with a *group name*, which also named their online group space. Entry into each group space was provided through the *group homepage* (see Figure 1). The group homepage showed the group's name (e.g., 'Labortech') and an *introductory description* at the top. It also identified the user (if logged in) and provided the user entry into his/her *member profile*, or a link for joining the group if the user was not a member.

---

[4] For an excellent discussion of the relationships between deliberative democracy and the design of groupware, see Noveck (2003).

[5] The version of Deme described below can be used at http://www.groupspace.org/wordpress/?page_id=54 and downloaded at http://www.groupspace.org/wordpress/?page_id=28.

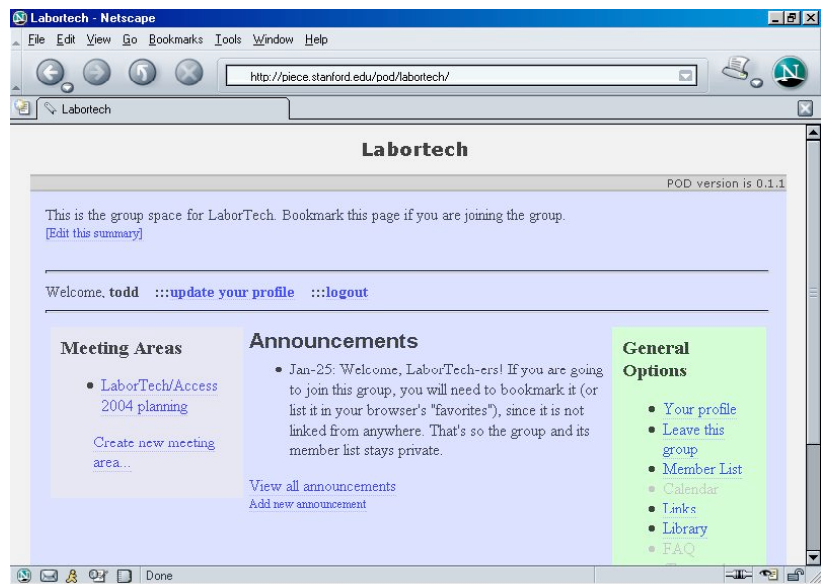

Figure 1. A Group Homepage from Deme (then 'POD') v0.1.1

These were familiar features to those who had used sites such as Yahoo! Groups. The somewhat novel feature of the group homepage was the availability of an arbitrary number of meeting areas. Each meeting area link took the user to a new page (a *meeting area viewer*), where group members could interact and/or deliberate. A meeting area might correspond to a committee or working group that was either a subgroup or a group connected to the group on whose homepage the meeting area was linked, or it might be set up around a topic for discussion or decision of interest to the group as a whole. A meeting area viewer is shown in Figure 2.

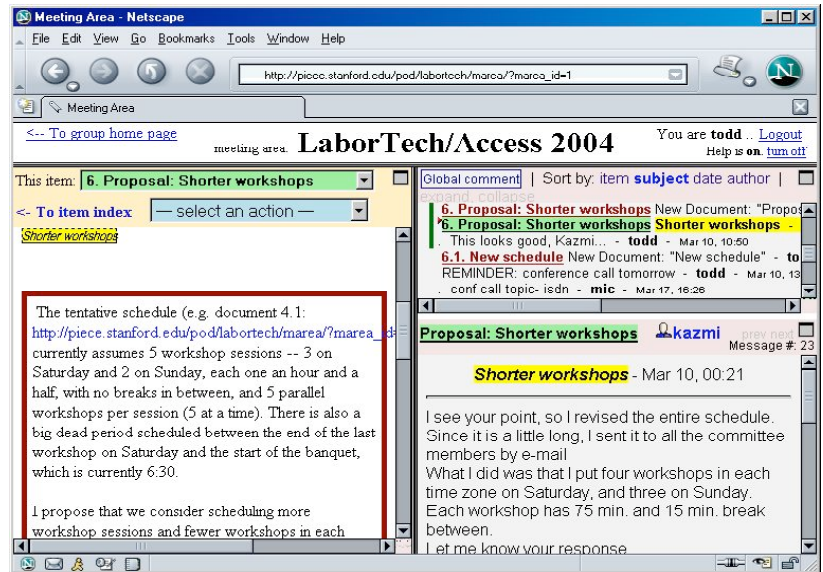

Figure 2. A meeting area from Deme v0.1.1

Beneath the *meeting area banner* at the top of the viewer page in Figure 2, the browser window was divided into three *panes*. Each pane could be viewed and operated upon either in the part of the screen shown (which was the *standard view* of a meeting area), or it could be made to fill most of the browser through the *enlarge button* located in the upper right corner of each pane. Under the banner, the standard view of a meeting area was divided vertically into the *discussion viewer* on the right side of the screen (consisting of a *comments index* pane that sat above a *comment reader* pane), and, on the left side of the screen, a pane known as the *folio viewer*. These left and right side viewers in the standard view were used to view and manipulate the two main types of objects in a meeting area: *items* and *comments*. Items comprised a meeting area's *folio*. Items were intended to be focuses of attention for the participants in a meeting area. The types of items included *documents, links, discussion items, nonbinding polls,* and *decisions*. Comments comprised a meeting area's *discussion*. A comment could be posted in reference to a particular item, or as a response to another comment, or as a *global comment*. In general, the meeting area was designed so that items were the objects of comments, and comments could refer to items. [6]

---

[6] This feature and some of the others described here persist in the latest version of Deme (Davies and Mintz 2009).

When a comment referred to an item, the *comment header*, as shown in both the comments index and comment reader, contained an *item reference*. Item references are shown as underlined red links at the beginning of the comment header in Figure 2. When an item reference was clicked on, it became *active* and the item to which its comment refers was loaded into the *item display*, which took up the bulk of the folio viewer and was located just beneath the *folio viewer control panel*. If an item reference was active, it was highlighted using both green shading and a small arrow in both the comments index and the comment reader. Clicking on a comment header, either via its item reference or via its *subject line,* made the comment itself (as opposed to the item reference) active. If there was an active comment, its subject line was highlighted in yellow in both the comments index and comment reader.

Comments could be viewed independently of the active item reference by clicking on their subject lines. But when an item reference was first clicked on, both the item reference and the comment that had been first associated with the item reference became active. Right after a click on an item reference, the referenced item was loaded into the item viewer so that the *comment reference* that was tied to the item reference could be seen in the display of the item, the comment was loaded into the discussion viewer, and the comment reference was highlighted in yellow inside the item display to indicate that both the comment and its associated item reference were active. A comment could reference an item either as a *general comment on the item* or as an *in-text comment*. In-text comments were unique to documents. The comment reference of an in-text comment could appear in any blank space within the document, and the document and the location of the comment reference together became the comment's item reference, indicating to Deme what the user should see in the folio viewer when an item reference was clicked on in the discussion viewer. All items could have general comments that referenced them.

As an example, in Figure 2 the user has clicked on the item reference '6. Proposal: Shorter Workshops', which is highlighted in green with a small arrow pointing to it in the comments index. This item reference appears in the comment header for the comment 'Shorter workshops', which was posted by 'kazmi'. The document itself appears on the left in the folio viewer, with the active comment reference highlighted in yellow above the text of the document. Documents could be entered directly (typed or pasted in as plain text), which allowed in-text commenting, or they could be uploaded in any format and made available for general comments.

There were additional features and subtleties in the design of the meeting area viewer. The main point was that the meeting area viewer was de-signed to embody the principles discussed in the section above on 'Comprehensive Deliberation'. Through a division between items and comments, and an architecture for referring to each, the meeting area viewer more closely approximated the processes of collaboration and item-centered discussion that happen in face-to-face meetings. Additional item types—discussion items, Web links, nonbinding polls, and decisions (e.g., majority, approval, plurality, and consensus procedures) were integrated into the meeting area to allow a full range of deliberation activities.

Consistent with our conclusions about how best to support groups, all versions of Deme have been free/open-source software (under the Affero General Public License) that can be accessed either through the server that we maintain or else installed on a group's own server.

## 6   Experience and Follow-up

An early release version of Deme was made available on Freshmeat.net in January of 2004, and group spaces were set up on Groupspace.org for tutoring new users, for internal development discussion, and for a group planning the LaborTech 2004 conference at Stanford. Over the following year, about a dozen groups used Deme on a regular basis, and many others created group spaces for trial. The response from users was generally positive, with many new users commenting that Deme had great potential to enhance participation in groups of which they were members. By 2005, however, the frame-based interface of the early Deme was behind the times, as frames gave way to more nimble user experiences based on AJAX. Deme has been rewritten a few times since the first version in order to keep up with advancing Web technology. But our experience with the first version taught us some enduring lessons for the design of an online deliberation platform:

- *Complexity demands visual guidance*. While users appreciated the functionality of Deme once they learned how it worked, the complexity of the early Deme interface proved too confusing for most new users, leading to lower levels of adoption and use. A redesigned interface (Figure 3) addressed this problem through affordances, icons, and labels (Davies et al. 2005; 2008).
- *Commenting must be integrated with email*. Because our first test group's Deme space was set up as a supplement to its regular email list, members continued to use the email list in addition to the group space, which caused confusion and duplicated effort. We concluded that Deme must offer to groups the ability to transfer their email list wholesale into Deme, with support for posting to (not just reading and being notified of) meeting area discussions via email. This feature was added in version 0.5 in July 2005.

- *Codebase must be built for incremental improvement*. Advances in Web technology and experiences with users dictated many changes to the software, but these proved difficult for new student programmers to implement in the PHP version. The advent of Web application frameworks such as Ruby on Rails and Django made possible a new way of designing complex websites that addresses this problem, and recent versions of Deme have been written in these frameworks.[7]

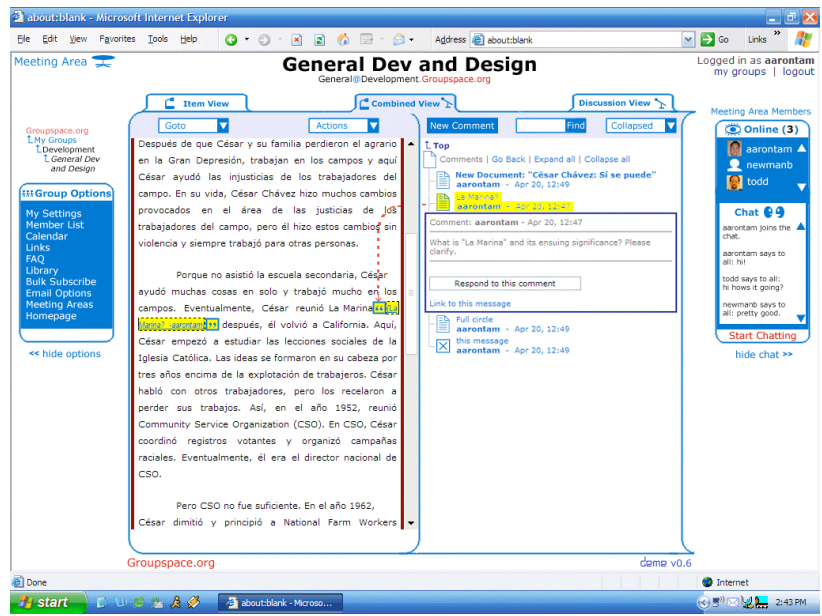

Figure 3. Meeting area redesign (mockup) from January 2006

## 7 Related Work

Many of Deme's features appeared in some form prior to the first version of Deme. Web-based tools existed for document-centered discussion (e.g., Quicktopic), collaborative authoring (e.g., TWiki), polling and integrating email with message boards (e.g., in Yahoo! Groups and phpBB), petition signing (e.g., PetitionOnline), survey design (e.g., Zoomerang), event scheduling (e.g., Meetup), and many other useful applications for groups.

---

[7] The latest version is at http://deme.stanford.edu (last accessed December 20, 2008).

Previous research prototypes had explored in-text comments of the type implemented in Deme (see Cadiz et al. 2000). Furthermore, interface designs had been developed to address the multiple points of focus that characterize group meetings; e.g., flexible split-screen interfaces in desktop applications such as the FreeAgent newsreader and the D3E discussion environment. We wanted to develop a platform that integrated many of these functional and interface ideas and was entirely Web-based, so that, ideally, a group's members could log into the platform from any computer on the Internet. Another project with goals broadly similar to ours has been the Communities of Practice Environment (CoPE) reported in Thaw, Feldman, and Li (2008) and Thaw et al. (forthcoming).

In the context of social science, our work generally aligns with the perspective known as 'deliberative democracy' (see Fishkin 2009; Gutmann and Thompson 1997), which holds that democracy can be enhanced by tying social decisions to thoughtful, fair, and informed dialogue among stakeholders, rather than through the filtering and manipulation of raw public opinion by power holders.

## 8  Conclusion

A common theme of participant-observations leading up to the design of Deme was that the need to make group decisions in face to face meetings often serves as an excuse for inner-circle, nontransparent decision making at many levels in society, ranging from small informal activist organizations to the U.S. Government. Our goal is to eliminate that excuse, so that stakeholders can legitimately demand to be included in decisions even if they cannot be present at face-to-face meetings or are not in an executive body. Our hope is that tools like Deme will eventually change the culture of democracy to one in which we expect more participatory inclusion from institutions and more participation from ourselves.

## 9  Acknowledgments

We are grateful for the assistance of many people who have contributed to Deme during the period reported in this chapter, including Mic Mylen, Rolando Zeledon, Bayle Shanks, Kent Koth, Art McGee, Sally Kiester, Laurence of Berkeley, and David Taylor.

This work was funded by a Public Scholarship Initiative grant from the Vice Provost for Undergraduate Education (VPUE) at Stanford administered by the Haas Center for Public Service, by a VPUE Departmental Grant to the Symbolic Systems Program, and an unrestricted gift to the Symbolic Systems Program by the late Ric Weiland, to whom we dedicate this chapter.